\documentclass[aps,prd,preprint,preprintnumbers,showpacs,nofootinbib,
superscriptaddress]{revtex4}
\usepackage{graphicx}
\usepackage{amsmath}
\usepackage{bbm}
\usepackage{bm}
\usepackage{epsfig}
\usepackage{amssymb}
\usepackage{slashed}
\usepackage{relsize}
\usepackage{color}
\usepackage{epsfig}
\usepackage{multirow}

\def\babar{\mbox{\slshape B\kern-0.1em{\smaller A}\kern-0.1em
    B\kern-0.1em{\smaller A\kern-0.2em R}}}

\begin{document}

\preprint{KIAS-PREPRINT-P12062}
\preprint{SLAC-PUB-15275}

\title{Resummation of relativistic corrections to exclusive productions of
charmonia in $\bm{e^+e^-}$ collisions
}

\author{Ying Fan}
\email[]{ying.physics.fan@gmail.com}
\affiliation{ Department of Physics, Korea University,
Seoul 136-701, Korea }

\author{Jungil Lee}
\email[]{jungil@korea.ac.kr}
\affiliation{ Department of Physics, Korea University,
Seoul 136-701, Korea }

\author{Chaehyun Yu}
\email[]{chyu@kias.re.kr}
\affiliation{School of Physics, KIAS, Seoul 130-722, Korea}
\affiliation{SLAC National Accelerator Laboratory,  \\
2575 Sand Hill Rd, Menlo Park, CA 94025, USA}

\begin{abstract}
We investigate two exclusive processes, $e^+ e^- \to \eta_c + \gamma$
and $e^+ e^- \to J/\psi + J/\psi$, at the center-of-momentum energy
$\sqrt{s}=10.58$ GeV within the framework of the nonrelativistic QCD
factorization approach.
A class of relativistic corrections is resummed to all orders in
the heavy-quark velocity $v$ and the corrections are large negative.
We further improve the prediction by including
available QCD next-to-leading-order corrections and the interference
between the QCD and relativistic corrections. The prediction for 
$\sigma[e^+ e^- \to \eta_c + \gamma]$ is about $50$\,fb.
In the case of $e^+ e^- \to J/\psi + J/\psi$ the standard 
nonrelativistic QCD prediction for the cross section is negative.
As an alternative, the vector-meson-dominance approach is employed
to compute the photon-fragmentation contribution of the process,
which gives the cross section $\sim 1$\,fb. This is an indication
that the uncalculated QCD higher-order corrections may be significant.
Our results can be tested against the forthcoming data from Belle II
and super $B$ factories.
\end{abstract}
\maketitle

\section{Introduction
\label{sec:intro}}
The nonrelativistic QCD (NRQCD) factorization approach \cite{NRQCD}
is a systematic theoretical tool to study the production and decay
of heavy quarkonia. In this approach, the production rate for a 
heavy quarkonium in a high-energy process is expressed as a linear
combination of NRQCD long-distance matrix elements (LDMEs) for
a heavy-quark-antiquark ($Q\bar{Q}_n$) pair into the physical 
quarkonium $H$. The velocity-scaling rules of NRQCD classify
the numerical importance of each LDME relative to the color-singlet
one in powers of $v$, the velocity of the quark and
antiquark in the $Q\bar{Q}_n$ rest frame.
The corresponding short-distance factors are calculable perturbatively. 
While the production of a quarkonium in hadron colliders
involves various channels with the $Q\bar{Q}_n$ pair whose quantum number
$n$ differs from that of $H$, an exclusive quarkonium production
process in $e^+e^-$ annihilation such as $e^+e^-\to J/\psi+\eta_c$
or $e^+e^-\to J/\psi+J/\psi$ is dominated by the color-singlet process
in which a produced $Q\bar{Q}$ pair has the same quantum number as 
that of a final-state quarkonium. This greatly reduces theoretical 
uncertainties. The factorization theorem for some exclusive quarkonium 
production processes in $e^+e^-$ annihilation is available \cite{fac},
although that for the inclusive quarkonium production is still a conjecture.

The only exclusive quarkonium production process that has been
observed at $B$ factories is $e^+ e^- \to J/\psi + \eta_c$ \cite{Abe:2002rb}.
Right after the first observation, it was revealed that the measured 
production rate \cite{Abe:2002rb} is greater than the leading-order (LO) 
NRQCD results \cite{Braaten:2002fi,Liu:2002wq} by an order of magnitude.
Although the subsequent measured values have 
decreased \cite{Abe:2004ww,Aubert:2005tj} in comparison with the first 
measurement, the discrepancy still remained. Through extensive theoretical
studies on the corrections 
at the next-to-leading order (NLO) in the strong coupling 
$\alpha_s$ \cite{Zhang:2005cha} and the resummed relativistic 
corrections \cite{Bodwin:2006dm,Bodwin:2006dn}, it was found that 
the interplay of the large QCD and relativistic corrections fill the gap
between the theory and data \cite{Bodwin:2007ga,Bodwin:2006ke,He:2007te}
within uncertainties. A recent result for the order-$\alpha_s v^2$
corrections also supports this conclusion~\cite{Dong:2012xx}.
However, some subtle issues that should be clarified still exist:
The current experimental data for $e^+ e^- \to J/\psi + \eta_c$
contain the events with a $\mu^+\mu^-$ pair plus at least two
charged tracks as the decay product of the $\eta_c$. If we take
into account the $\eta_c$ decay modes without charged tracks,
then the actual cross section can be larger than the current data.
While the size of the relativistic corrections has been estimated rather
precisely with the aid of the resummation technique, the QCD 
corrections were computed only at order $\alpha_s$. The amount
of higher-order (HO) QCD corrections are not estimated yet.

Another interesting issue is regarding the color-singlet LDMEs.
The color-singlet LDMEs are usually determined by comparing 
experimental data with theoretical predictions for the 
electromagnetic decays of the $S$-wave charmonia, 
$J/\psi \to l^+l^-$ and $\eta_c \to \gamma \gamma$. It is known 
that the relativistic corrections to the charmonium decay rates
are significant. In order to determine the color-singlet LDMEs
including the relativistic corrections to these electromagnetic
decays, we need at least two constraints for each decay rate.
One is the experimental decay rate. As for the other constraint,
one may choose the generalized Gremm-Kapustin 
relation~\cite{Bodwin:2006dn} that relates the LDMEs of LO and 
HO~\cite{Bodwin:2007fz,Bodwin:2008vp,Chung:2010vz,Guo:2011tz}. 
As an alternative, one can use the measured rate 
$\Gamma[J/\psi\to\textrm{LH}]$ for the $J/\psi$ decay
into light hadrons (LH) \cite{He:2007te}. However, the 
numerical values for the LDMEs have a strong dependence on the choice:
The LO LDME for $J/\psi$ at $m_c=1.4$\,GeV is 
$0.440$\,GeV$^3$~\cite{Bodwin:2007fz} and 
$0.573$\,GeV$^3$~\cite{He:2007te}. The corresponding values
for $v^2$ are $0.225$~\cite{Bodwin:2007fz} and $0.089$~\cite{He:2007te}.
Therefore, the phenomenological study on the exclusive quarkonium
production rate in $e^+e^-$ annihilation may provide us with a 
good chance of improving the accuracies in the determination of
the color-singlet LDMEs. An accurate determination of the 
color-singlet LDMEs may lead to improving the determinations
of various color-octet LDMEs that are involved in the inclusive
productions of $J/\psi$~\cite{co}.

One may also consider exclusive $S$-wave quarkonium processes in 
$e^+e^-$ annihilation other than the observed one
$e^+ e^- \rightarrow J/\psi + \eta_c$. 
In Refs. \cite{Bodwin:2002fk,Bodwin:2002kk},
the process $e^+ e^- \rightarrow J/\psi + J/\psi$ was first 
introduced as a possible source of contaminating the data samples for
$e^+ e^- \rightarrow J/\psi + \eta_c$. The angular-distribution
analysis of the $e^+ e^- \rightarrow J/\psi + \eta_c$ events at
Belle showed that there were no such contaminations. In addition,
the Belle Collaboration has not observed the signals for 
$e^+ e^- \rightarrow J/\psi + J/\psi$ yet and only an upper
bound of the cross section has been reported:
$\sigma[e^+ e^- \rightarrow J/\psi + J/\psi]%
 \times \mathcal{B}_{>2}[J/\psi]%
 <  9.1$ fb \cite{Abe:2004ww}.
According to the QCD NLO correction in Ref. \cite{Gong:2008ce},
the NLO corrected cross section is significantly smaller than the 
LO prediction. As is stated in Refs. \cite{Bodwin:2002fk,Bodwin:2002kk},
the process is dominated by the photon fragmentation.
The relativistic and QCD corrections to this contribution can be
summed effectively by making use of the vector-meson-dominance (VMD)
approach \cite{Bodwin:2006yd}. The cross section within the VMD 
approach is 
$\sigma[e^+ e^- \to J/\psi+J/\psi]=1.69\pm 0.35$\,fb at the 
center-of-momentum (CM) energy 
$\sqrt{s}=10.58$\,GeV \cite{Bodwin:2006yd}, which is significantly
smaller than the LO prediction and is consistent with the 
nonobservation of the process at Belle. The production of a 
charge-conjugation parity $+1$ quarkonium associated with a photon, 
$e^+ e^- \rightarrow H+\gamma$, was first suggested in 
Ref. \cite{Chung:2008km} as a nice probe to the color-singlet 
mechanism of NRQCD and the convergence of relativistic corrections,
especially for $\eta_c(2S)$. Later, both the QCD NLO and 
relativistic corrections to this process were calculated and
found to be considerable~\cite{Sang:2009jc,Li:2009ki}.

In this work, we investigate the resummed relativistic correction
and its interplay with the QCD NLO correction to the exclusive
processes 
$e^+ e^- \to \eta_c + \gamma$ and $e^+ e^- \to J/\psi + J/\psi$ at 
 $\sqrt{s}=10.58$\,GeV. Because the QCD NLO and 
relativistic corrections to $e^+ e^- \to J/\psi + \eta_c$ are large, 
one may guess that the two processes listed above may also 
acquire significant corrections relative to the LO contribution.
Our calculation reveals that the corrections are indeed large negative.
The resummation of relativistic corrections enhances the cross 
section for $e^+ e^- \to \eta_c + \gamma$ in comparison with the
order-$v^2$ relativistic correction. In the case of 
$e^+ e^- \to J/\psi + J/\psi$, the QCD NLO and relativistic
corrections resummed to all orders in $v$ 
make the cross section negative
within the standard NRQCD factorization approach. By employing the 
VMD approach to compute the photon-fragmentation contribution,
we cure the problem. The result shows that the QCD HO corrections
may have significant contributions to $e^+ e^- \to J/\psi + J/\psi$.
Therefore, these two exclusive processes may provide us with
independent phenomenological constraints to the color-singlet 
NRQCD LDMEs for the $S$-wave charmonia. 

The remainder of this paper is organized as follows.
In Sec. \ref{sec:method}, we briefly describe the strategy of
resumming relativistic corrections in a quarkonium process.
In Sec.~\ref{sec:etac}, we compute the cross section for 
$e^+ e^- \to \eta_c + \gamma$, in which 
the QCD NLO and resummed relativistic corrections, and their 
interference are included within the standard NRQCD factorization
approach. The corresponding cross section for $e^+ e^- \to J/\psi+J/\psi$ 
is given in Sec.~\ref{sec:jpsi}. In the case of $e^+ e^- \to J/\psi+J/\psi$,
that is dominated by the photon fragmentation, the result within the
standard NRQCD approach gives negative cross section. As an alternative,
we also employ the VMD approach to find an improved prediction.
Finally, we conclude in Sec.~\ref{sec:con}.
\section{Strategy of resumming relativistic corrections
\label{sec:method}}
In this section, we list the NRQCD factorization formula for 
the cross sections of
$e^+e^-\to \eta_c +\gamma$ and $e^+ e^- \rightarrow J/\psi + J/\psi$,
and summarize the strategy of resumming relativistic corrections.
\subsection{NRQCD factorization formula}
According to the NRQCD factorization formalism \cite{NRQCD}, the cross 
sections for the exclusive processes
$e^+e^-\to \eta_c +\gamma$ and $e^+ e^- \rightarrow J/\psi + J/\psi$
can be expressed as:
\begin{subequations}
\label{fac-sigma}
\begin{eqnarray}
\sigma [e^+e^-\to\eta_c+\gamma]
&=& \sum_n\frac{F_n}{m_c^5}
\langle  \mathcal{O}_n^{\eta_c}  \rangle,
\\
\sigma [e^+e^-\to J/\psi+J/\psi] &=& \sum_{m,n}\frac{F_{m,n}}{m_c^8}
\langle  \mathcal{O}_m^{J/\psi}  \rangle
\langle  \mathcal{O}_n^{J/\psi}  \rangle,
\end{eqnarray}
\end{subequations}
where $\langle  \mathcal{O}_n^{H}\rangle$ is the NRQCD LDME for
production of a quarkonium $H$ from a $c\bar{c}_n$ pair with
a spectroscopic state $n$, $F_n$ and $F_{m,n}$ are dimensionless
perturbative short-distance coefficients that are independent
of the long-distance nature of $H$, and $m_c$ is the charm-quark mass.

In general, the quantum number for $c\bar{c}_n$ does not
have to be the same as that for the quarkonium $H$ and the numerical
importance of a $c\bar{c}_n$ channel relative to the color-singlet
channel is classified in powers of $v$ in the velocity-scaling
rules of NRQCD. In the exclusive process (\ref{fac-sigma}),
$c\bar{c}_n$ channels with the quantum number identical to that of
$H$ contribute dominantly. Thus the series (\ref{fac-sigma}) can be
well approximated by the relativistic corrections and the index $n$
can be understood as the power in $v^2$ relative to the LO
color-singlet NRQCD LDME $\langle  \mathcal{O}_0^{H}\rangle$.
The expression (\ref{fac-sigma}) can further be simplified once
we apply the vacuum-saturation approximation to express the 
production LDME in terms of the corresponding decay LDME 
$\langle  \mathcal{O}_n  \rangle_{H}$:
$\langle  \mathcal{O}_n^{H}  \rangle \approx%
(2J+1) \langle  \mathcal{O}_n  \rangle_{H}$, where
$J$ is the total-angular-momentum quantum number of $H$.
In the Coulomb gauge, the relative-order-$v^{2n}$ decay LDMEs for
$H=J/\psi$ and $\eta_c$ are expressed as
\begin{subequations}
\label{ME-HO}
\begin{eqnarray}
\langle  \mathcal{O}_n  \rangle_{J/\psi} &=&
\langle J/\psi(\lambda) |
\psi^\dagger \left(-\tfrac{i}{2}\tensor{\bm{D}}\right)^{2a}
\bm{\sigma}\cdot\bm{\epsilon}^{\ast}(\lambda)\chi|0\rangle
\langle 0 |
\chi^\dagger \left(-\tfrac{i}{2}\tensor{\bm{D}}\right)^{2b}
\bm{\sigma}\cdot\bm{\epsilon}(\lambda)\psi|J/\psi(\lambda)\rangle,
\nonumber\\
\\
\langle  \mathcal{O}_n  \rangle_{\eta_c} &=&
\langle \eta_c |
\psi^\dagger \left(-\tfrac{i}{2}\tensor{\bm{D}}\right)^{2a}
\chi|0\rangle
\langle 0 |
\chi^\dagger \left(-\tfrac{i}{2}\tensor{\bm{D}}\right)^{2b}
\psi|\eta_c\rangle,
\end{eqnarray}
\end{subequations}
where $n=a+b$,\footnote{If $a\neq b$, the right side of
Eq.~(\ref{ME-HO}) is understood to be the average with 
its Hermitian conjugate.}
$\psi^\dagger$ and $\chi$ are two-component Pauli spinor fields  that
create a heavy quark and a heavy antiquark, respectively,
$\sigma^i$ is a Pauli matrix, $\bm{D}$
is the spatial component of the covariant derivative, and
$\lambda$ and $\bm{\epsilon}(\lambda)$ are the helicity and polarization
vector of $J/\psi$, respectively.\footnote{
In general, there are additional contributions that depend on
chromoelectromagnetic field operators that appear from relative order $v^4$.
At the relative order $v^4$ such contribution can be expressed in terms of
the operators of the same relative order in
Eq.~(\ref{ME-HO}) \cite{Bodwin:2002hg,Bodwin:2012xc}.
In this work, we neglect the gauge-field contribution in the Coulomb gauge
and, therefore, the contributions of the chromoelectromagnetic field operators
are also neglected.}
The state $|H\rangle$ is normalized nonrelativistically:
$\langle H(\bm{P})|H(\bm{P}')\rangle=(2\pi)^3\delta^{(3)}(\bm{P}-\bm{P}')$.
The LO color-singlet LDMEs are given by
\begin{subequations}
\label{ME-LO}
\begin{eqnarray}
\langle O_0 \rangle_{J/\psi}  &=&
\left|\langle 0 |
\chi^\dagger
\bm{\sigma}\cdot\bm{\epsilon}(\lambda)\psi|J/\psi(\lambda)\rangle
\right|^2,
\\
\langle O_0 \rangle_{\eta_c}  &=&
\left|\langle 0 |
\chi^\dagger
\psi|\eta_c\rangle
\right|^2.
\end{eqnarray}
\end{subequations}
\subsection{Resummation of relativistic corrections\label{subsec:resum}}
The amplitude of an $S$-wave charmonium $H$ produced associated 
with a particle $a$ in an $e^+e^-$ annihilation can be expressed as
\begin{equation}
\label{expansion}%
\mathcal{A}[e^+e^-\to H+a]=\sum_{n=0}^\infty
\left.\left[\frac{1}{n!}\left(\frac{\partial}
{\partial \bm{q}^2}\right)^n \mathcal{M}({\bm{q}^2})\right]
\right|_{\bm{q}^2=0}
\langle \bm{q}^{2n} \rangle_{H}
\langle \mathcal{O}_0 \rangle_{H}^{1/2},
\end{equation}
where $\mathcal{M}({\bm{q}^2})$ is the corresponding parton-level amplitude
with the standard normalization in the NRQCD factorization approach
in which the angular dependence of half the relative momentum $\bm{q}$
of the $c$ and $\bar c$ has been averaged in the $c\bar{c}$ rest frame.
$\langle\bm{q}^{2n}\rangle_{H}$ is the ratio of the HO LDME of 
order $v^{2n}$ relative to the LO one. For $H=J/\psi$,
\begin{equation}
\langle\bm{q}^{2n}\rangle_{J/\psi} = \frac{\langle 0|
\chi^\dagger (-\frac{i}{2}\overleftrightarrow{\bm{D}})^{2n}
\bm{\sigma}\cdot\bm{\epsilon}(\lambda) \psi | J/\psi(\lambda) \rangle}
{\langle 0|\chi^\dagger\bm{\sigma}\cdot\bm{\epsilon}
(\lambda) \psi |J/\psi(\lambda)  \rangle}.
\end{equation}
Eventually in the factorization formula (\ref{fac-sigma}), the long-distance
factors in Eq.~(\ref{expansion}) are expressed in terms of the LDMEs in
Eq.~(\ref{ME-HO}) that  are, in principle,
independent of the LO LDMEs $\langle O_0 \rangle_{H}$ in Eq.~(\ref{ME-LO}).
In Ref.~\cite{Bodwin:2006dn}, the Cornell potential model was used to
express the ratio $\langle\bm{q}^{2n}\rangle_{H}$ in terms of
$\langle\bm{q}^{2}\rangle_{H}$:
\begin{equation}
\label{G-GK}
\langle\bm{q}^{2n}\rangle_{H} =
\langle\bm{q}^{2}\rangle_{H}^{n},
\end{equation}
that we call the generalized Gremm-Kapustin
relation\footnote{The original form of the Gremm-Kapustin relation is
$\langle \bm{q}^2 \rangle_{H} \approx (m_{H} - 2 m_c) m_c$
\cite{Gremm:1997dq}, where $m_H$ is the $S$-wave quarkonium mass.}.
The relation (\ref{G-GK}) neglects the gauge-field contribution to the
covariant derivative in the Coulomb gauge and neglects the spin-flipping
interactions so that
$\langle\bm{q}^{2n}\rangle_{\eta_c}\approx%
\langle\bm{q}^{2n}\rangle_{J/\psi}$. Various applications and detailed
descriptions of the uncertainties of applying the relation can be found 
in Refs.~\cite{Bodwin:2006dn,Bodwin:2007fz,%
Bodwin:2007ga,Bodwin:2008vp,Bodwin:2012xc}.

The generalized Gremm-Kapustin relation (\ref{G-GK}) allows one to 
resum a class of relativistic corrections to all orders in $v$.
The resultant amplitude is
\begin{equation}
\label{eq:resum}%
\mathcal{A}[e^+e^-\to H+a]=
\mathcal{M}(\langle \bm{q}^2\rangle_{H})
\langle \mathcal{O}_0 \rangle_{H}^{1/2}\;.
\end{equation}
It is straightforward to apply this resummation method to
$e^+e^-\to \eta_c+\gamma$ and $e^+e^-\to J/\psi+J/\psi$.
\subsection{The VMD approach}
In Ref.~\cite{Braaten:2002fi}, the importance of the 
photon-fragmentation contribution in the $J/\psi$ production
in $e^+e^-$ annihilation was considered and the competition
between the QED and QCD contributions was first observed.
The dominance of the photon fragmentation
in $e^+e^-\to J/\psi+J/\psi$
at the $B$ factories was predicted in
Refs.~\cite{Bodwin:2002fk,Bodwin:2002kk,Bodwin:2006yd}.
In spite of a suppression factor $\alpha^2/\alpha_s^2$ relative to
the QCD subprocesses of other exclusive two-quarkonium production
in $e^+e^-$ annihilation, the photon-fragmentation contribution is
largely enhanced by a kinematic factor $[\sqrt{s}/(2m_c)]^4$ that is associated 
with the fragmentation of a photon into a $c\bar{c}$ pair.
The rate is further enhanced by a factor of $\log [\sqrt{s}/(2m_c)]$
originated from the collinear emission of a virtual photon 
in the forward region.
Similar enhancement of the photon-fragmentation contribution was predicted
in inclusive $J/\psi$ production with sufficiently large transverse momentum
$p_T$ at hadron colliders \cite{He:2009cq,Fan:2012vw}.

As is stated in Refs.~\cite{Bodwin:2006yd,Bodwin:2007ga}, the prediction of
the photon-fragmentation contribution in $e^+ e^- \to J/\psi + J/\psi$
can be improved by employing the VMD approach. We shall find
later that after combining the QCD NLO and resummed relativistic corrections to
$e^+ e^- \to J/\psi + J/\psi$, the cross section becomes negative
within the standard NRQCD factorization approach. As an alternative,
we employ the VMD approach to cure this symptom.

If the photon-fragmentation dominance is valid, then one can 
neglect the non-fragmentation contributions that make a  
gauge-invariant subset. In that case, the amplitude is factorized into 
the product of 
$\mathcal{M}[e^+e^-\to \gamma^*+\gamma^*]$ and the $\gamma^*\to J/\psi$
fragmentation factor that involves the local electromagnetic current
\begin{equation}\label{curr}%
    J^\mu(x)=e_c e \bar{c}(x)\gamma^\mu c(x)
\end{equation}
for a free $c\bar{c}$ pair created at a fixed point $x$.
Here, $e$ is the electron charge and $e_c$ is the fractional
electric charge of the charm quark.
 In VMD, the long-distance
factor for the transition of the pair to $J/\psi$ is expressed as
\begin{equation}\label{eq:ME-g}
    \langle J/\psi(\lambda)| J^\mu (x=0) |0\rangle =
    g_{J/\psi\gamma} \epsilon^{\ast\mu}(\lambda),
\end{equation}
where $g_{J/\psi\gamma}$ is the effective $J/\psi$-photon coupling.
Applying this vertex to $J/\psi \to e^+ e^-$, one can determine the coupling 
$g_{J/\psi \gamma}$ \cite{Bodwin:2006yd,Bodwin:2007ga} as
\begin{equation}\label{eq:gVgamma}%
    g_{{J/\psi}\gamma} = \left( \frac{3 m_{J/\psi}^3}{4\pi\alpha^2}
    \Gamma[J/\psi\rightarrow e^+e^-] \right)^{\frac{1}{2}}\; .
\end{equation}
If we can neglect the virtual-gluon corrections to
the photon-fragmentation process that connect the two $c\bar{c}$
lines, then the vertex (\ref{eq:ME-g}) collects
the QCD HO and relativistic corrections effectively.
We can make an educated guess that a virtual-gluon correction, that appears
from relative order $\alpha_s^2$, is negligible because the contribution is
suppressed by $\alpha_s^2$ relative to the LO and the kinematic enhancement is
reduced due to the stronger virtualites of the internal lines.
Because we use the measured values for the decay rate
$\Gamma[J/\psi\to e^+ e^-]$ and the $J/\psi$ mass $m_{J/\psi}$, 
in evaluating the effective coupling (\ref{eq:gVgamma}), large theoretical 
uncertainties can be replaced with tiny experimental errors.
We take the scale $m_{J/\psi}$, which is the momentum scale at the vertex,
in computing the fine structure constant $\alpha$ in Eq.~(\ref{eq:gVgamma}).
\subsection{Interference\label{inter}}
In a previous analysis in Ref.~\cite{Bodwin:2007ga} on
$e^+ e^- \to J/\psi + \eta_c$, the authors found that both QCD NLO and
resummed relativistic corrections to the cross section were significant.
It was also reported that the interference between
the two corrections was about 26\,\% of the LO cross section.
Thus we expect that the interference term may also be important
in other processes that have large QCD and relativistic corrections.
In addition to this interference there is another
order-$\alpha_s v^2$ correction that can be computed directly from
the order-$\alpha_s$ diagrams by keeping the $v$ dependence.
Such order-$\alpha_s v^2$ corrections were computed for
$J/\psi \to \ell^+ \ell^-$ \cite{Bodwin:2008vp},
$B_c \to \ell \nu$ \cite{Lee:2010ts},
$\eta_c \to \gamma \gamma$ and LH \cite{Jia:2011ah,Guo:2011tz},
$e^+ e^- \to J/\psi + \eta_c$ \cite{Dong:2012xx},
$h_c$, $h_b$ and $\eta_b \to$ LH \cite{Li:2012rn},
and $J/\psi\to 3\gamma$ \cite{Feng:2012by}.
Many of the corrections are tiny, for example, at most $0.3$\,\%
in $J/\psi \to \ell^+ \ell^-$ \cite{Bodwin:2008vp}.
Some exceptions are 
$h_c \to$ LH \cite{Li:2012rn} and 
$J/\psi\to 3\gamma$ \cite{Feng:2012by},
which have significant order-$\alpha_s v^2$ corrections.

In this work, we improve the theoretical prediction for the production
rates of $e^+e^-\to \eta_c+\gamma$ and $e^+e^-\to J/\psi+J/\psi$ by including
the order-$\alpha_s$ and order-$v^\infty$ corrections, and their 
interference.\footnote{In the remainder of this paper, we use the
notation $v^\infty$ in order to indicate that a quantity is resummed
to all orders in $v$.}
The computation of the pure order-$\alpha_s v^2$ contribution, that is
out of the scope of this work, is not included.
The missing contribution is of the same order as the interference term
that may bring in theoretical errors.
Equation (42) of Ref.~\cite{Bodwin:2007ga} includes the three contributions
listed above to the total cross section $\sigma_{\rm tot}$:
\begin{equation}\label{interference}
    \sigma_{\rm tot}=
    \sigma_{v^\infty}+\sqrt{\sigma_{v^\infty}}
    \frac{\sigma_{\rm NLO}-\sigma_0}{\sqrt{\sigma_0}}
= \sigma_{v^\infty} 
+ \sqrt{\sigma_{v^\infty}\sigma_0} \left( K_{\alpha_s}-1\right),
\end{equation}
where $\sigma_0$, $\sigma_{v^\infty}$, and $\sigma_{\rm NLO}$ are 
the cross sections at the LO, order $v^\infty$, and order $\alpha_s$,
respectively, and $K_{\alpha_s}\equiv\sigma_{\rm NLO}/\sigma_0$.
The approximation for the interference term in Eq. (\ref{interference})
is constructed under the assumption that $K_{\alpha_s}$ and $v^\infty$
corrections are independent of the polarization $\lambda$ of $H$.
Violation of this assumption may cause additional theoretical errors.
\section{$\bm{e^+e^-\to\eta_c+\gamma}$ \label{sec:etac}}
\begin{figure}[t]
\begin{center}
{\epsfig{figure=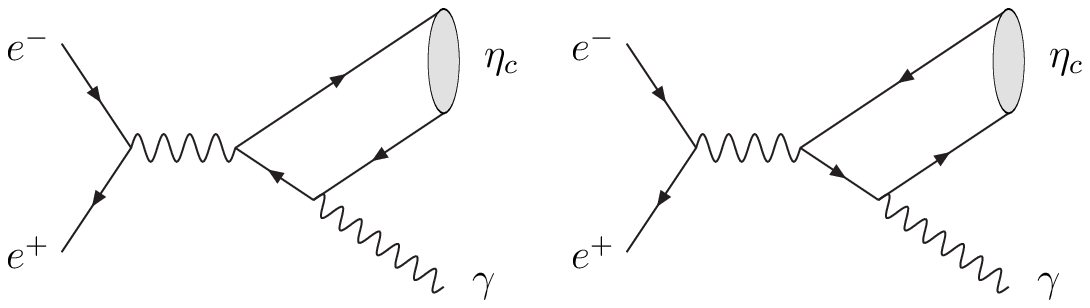,width=0.7\textwidth}}
\end{center}
\vspace{-1.0cm}
\caption{Feynman diagrams for $e^+ e^- \to \eta_c + \gamma$ at LO in $\alpha_s$.
}
\label{fig:diag1}
\end{figure}
In this section, we compute the cross section for $e^+ e^- \to \eta_c + \gamma$
at $\sqrt{s}=10.58$\,GeV by employing the strategies given in
Sec.~\ref{sec:method}. In $e^+e^-$ annihilation, this process proceeds with
the parton process $e^+e^-\to c\bar{c}({}^1S_0^{[1]}) + \gamma$,
where $c\bar{c}({}^1S_0^{[1]})$ is the color-singlet spin-singlet $S$-wave
$c\bar{c}$ pair. Because both QED and QCD have
the charge-conjugation symmetry and $\eta_c$ is an even charge-conjugation
eigenstate, the process factors into the leptonic part $e^+e^-\to \gamma^*$
and the hadronic part $\gamma^*\to c\bar{c}({}^1S_0^{[1]})+\gamma$.
At the LO in $\alpha_s$, two Feynman diagrams contribute to this process
as shown in Fig. \ref{fig:diag1}.
\subsection{Amplitude}
In order to construct the amplitude for $e^+e^-\to c\bar{c}({}^1S_0^{[1]}) + \gamma$,
we first write the amplitude $\mathcal{M}$ for the process
$e^+(k_2)e^-(k_1)\to c(p)\bar{c}(\bar{p}) + \gamma(P_\gamma)$ as
\begin{equation}\label{eq:amplitude}
    \mathcal{M}=L^\mu \mathcal{H}_\mu,
\end{equation}
where the momentum of each particle is written in the parentheses following
each particle identity. Here, the leptonic current $L^\mu$ is defined by
\begin{equation}\label{eq:Lmu}
    L^\mu=\frac{e}{s}\bar{v}(k_2)\gamma^\mu u(k_1).
\end{equation}
The hadronic current $\mathcal{H}_\mu$ is given by
$\mathcal{H}_\mu = \bar{u}(c) \mathcal{C}_\mu v(\bar{c})$
with $\mathcal{C}_\mu=\mathcal{C}_{\mu\nu} \epsilon^{\ast\nu}$,
$\epsilon$ is the polarization vector of the photon,
and the Dirac and color indices of $\mathcal{C}_{\mu\nu}$ are suppressed.
The momenta $p$ and $\bar{p}$ for the on-shell $c$ and $\bar{c}$, respectively,
are linear combinations of the quarkonium momentum $P$ and half the relative
momentum $q$:
\begin{subequations}
\begin{eqnarray}
p &=& \tfrac{1}{2}P+q, \\
\bar{p} &=& \tfrac{1}{2}P-q.
\end{eqnarray}
\end{subequations}
At the $c\bar{c}$ rest frame, $P=[2E(q),\bm{0}]$, $q=(0,\bm{q})$,
$p=[E(q),\bm{q}]$, and $\bar{p}=[E(q),-\bm{q}]$,
where $E(q)=\sqrt{m_c^2+\bm{q}^2}$.

The color-singlet spin-singlet contribution
can be projected out from $\mathcal{H}_\mu$ by taking the
trace after multiplying the projectors for the spin-singlet [$\Pi_1(p,\bar{p})$]
and the color-singlet ($\pi_1$) state. The corresponding projectors are defined by
\begin{subequations}
\begin{eqnarray}
\label{eq:PI-1}
\Pi_1(p,\bar{p}) &=&
\frac{1}{4\sqrt{2}E(q)[E(q)+m_c]}
(\bar{p}\!\!\!/-m_c)\gamma^5 [P\!\!\!\!/+2 E(q)](p\!\!\!/+m_c),
\\
\pi_1 &=& \frac{1}{\sqrt{N_c}} \mathbbm{1},
\end{eqnarray}
\end{subequations}
where $N_c=3$ is the number of colors and $\mathbbm{1}$ is the unit color matrix.
The spin projector (\ref{eq:PI-1}) contains the $q$ dependence to all orders
in $v$~\cite{Bodwin:2002hg}.
Then the $c\bar{c}({}^1S_0^{[1]})$ contribution $\mathcal{H}_\mu[c\bar{c}({}^1S_0^{[1]})]$
to the hadronic current is finally obtained by averaging over the directions
of $\bm{q}$ in the $c\bar{c}$ rest frame:
\begin{equation}\label{eq:angave1}
    \mathcal{H}_\mu[c\bar{c}({}^1S_0^{[1]})]=
   \int \frac{d\Omega_{\bm{q}}}{4\pi}\, \textrm{Tr}
   \left[\mathcal{H}_\mu \Pi_1 (p,\bar{p})\otimes \pi_1\right]\,,
\end{equation}
where $\Omega_{\bm{q}}$ is the solid angle of $\bm{q}$
and the trace is over the color and spinor indices.
Now the complete $q$ dependence in Eq.~(\ref{eq:angave1}) is
a function of $\bm{q}^2$. Carrying out the standard perturbative matching
of the full-QCD amplitude for the hadronic current to the NRQCD 
counterpart and employing the resummation strategy stated 
in Sec.~\ref{subsec:resum}, we obtain the amplitude for 
$e^+e^-\to\eta_c+\gamma$ as
\begin{equation}
\label{eq:Aetac}%
\mathcal{A}[e^+e^-\to\eta_c+\gamma] =
L_\mu
\left.
\left[\frac{1}{2N_c \sqrt{E(q)}}
\mathcal{H}_\mu[c\bar{c}({}^1S_0^{[1]})]
\right]
\right|_{\bm{q}^2=\langle \bm{q}^2 \rangle_{\eta_c}}
\langle O_{0} \rangle_{\eta_c}^{1/2}.
\end{equation}
\subsection{Kinematics}
Because of the average over the angle for $\bm{q}$ that is required to project
out the $S$-wave state, additional consideration of the $\bm{q}$ dependence
in the $c\bar{c}$ amplitude is needed. As is stated in Ref.~\cite{Bodwin:2007ga},
it is convenient to choose the $e^+e^-$ CM frame in which the final momenta are along
the $z$-axis. In this CM frame, the momenta for
$e^-$, $e^+$, $\eta_c$, and $\gamma$ are given by
\begin{subequations}
\begin{eqnarray}
  k_1 &=& \frac{\sqrt{s}}{2}
          \left(1, \sin\theta, 0, \cos\theta\right), \\
  k_2 &=& \frac{\sqrt{s}}{2}\left(1, -\sin\theta, 0, -\cos\theta\right),  \\
  P &=& \left(E_P, 0, 0, P_\textrm{CM}\right) ,\\
  P_\gamma &=& \left(P_\textrm{CM}, 0, 0,-P_\textrm{CM}\right) ,
\end{eqnarray}
\end{subequations}
where $\theta$ is the scattering angle, $E_P=[s+4 E^2(q)]/(2\sqrt{s})$, and
$P_\textrm{CM}=[s-4 E^2(q)]/(2\sqrt{s})$.
The explicit form of $q$ in this CM frame is given by
\begin{equation}\label{}
q = |\bm{q}|\left(
\gamma_q\beta_q\cos\theta_q, \sin\theta_q\cos\phi_q,
\sin\theta_q\sin\phi_q, \gamma_q\cos\theta_q \right),
\end{equation}
where $|\bm{q}|\equiv\sqrt{-q^2}$ and the two factors
$\gamma_q = E_P/[2 E(q)]$ and $\beta_q = P_\textrm{CM}/E_P$ involve
the boost from the $c\bar{c}$ rest frame to this $e^+e^-$ CM frame.
$\theta_q$ and $\phi_q$ are the polar and azimuthal 
angles, respectively, of $q^*$ that is the explicit form 
of $q$ in the $c\bar{c}$ rest frame:
\begin{equation}\label{qrest}
    q^\ast = |\bm{q}|\left( 0, \sin\theta_q\cos\phi_q,
     \sin\theta_q\sin\phi_q, \cos\theta_q \right).
\end{equation}

\begin{table}[t]
\caption{\label{table1}%
The cross sections $\sigma_0$, $\sigma_{v^\infty}$, and
$\sigma_\textrm{tot}$ for $e^+e^-\rightarrow \eta_c+\gamma$
in units of fb at the scales $\mu=m_c$ and $2 m_c$
with the input parameters $m_c$, $\langle O_{0} \rangle_{\eta_c}$,
$\langle \bm{q}^2 \rangle_{\eta_c}$, in units of GeV, GeV$^3$,
and GeV$^2$, respectively. The errors of the cross sections 
reflect the uncertainties of the NRQCD LDMEs only.
}
\begin{ruledtabular}
\begin{tabular}{ccccccc}
\multirow{2}{*}{$m_c$}
&
\multirow{2}{*}{$\langle O_{0} \rangle_{\eta_c}$}
&
\multirow{2}{*}{$\langle \bm{q}^2 \rangle_{\eta_c}$}
&
\multirow{2}{*}{$\sigma_0$}
&
\multirow{2}{*}{$\sigma_{v^\infty}$}
&
\multicolumn{2}{c}{$\sigma_\textrm{tot}$}
\\
&
&
&
&
&
$\mu=m_c$
&
$\mu=2m_c$
\\
\hline
  $1.4$
&
$0.437^{+0.111}_{-0.105}$
&
$0.442\pm 0.143$
&
$83.2^{+21.1}_{-20.0}$
&
$68.9^{+18.0}_{-17.0}$
&
$51.5^{+13.6}_{-12.8}$
&
$55.3^{+14.5}_{-13.7}$
\end{tabular}
\end{ruledtabular}
\end{table}
\subsection{Cross section}
Summing over the final spins and averaging over the initial spins
of the absolute square of the amplitude
$\mathcal{A}[e^+e^-\to\eta_c+\gamma]$ in Eq.~(\ref{eq:Aetac}),
we obtain the spin-averaged squared amplitude
$\overline{|\mathcal{A}[e^+e^-\to\eta_c+\gamma]|^2}$.
Dividing it by the flux factor $2s$, and integrating over the phase space,
we find the cross section.

In order to evaluate the cross section for $e^+e^-\to\eta_c+\gamma$
we first choose the numerical values for the input parameters.
We choose $\alpha=1/131$ that is the running coupling constant evaluated
at scale $\mu=\sqrt{s}$~\cite{Bodwin:2007fz,Chung:2008km}.
As is shown in the first three columns
in Table \ref{table1}, we choose $m_c$, $\langle O_{0} \rangle_{\eta_c}$,
and $\langle \bm{q}^2 \rangle_{\eta_c}$ from 
Refs.~\cite{Bodwin:2007fz,Chung:2008km}. The LO cross section
$\sigma_0$ is consistent with previous results in 
Refs. \cite{Chung:2008km,Sang:2009jc} \footnote{
The errors of the cross sections reflect the 
uncertainties of the NRQCD LDMEs only.}.
We recall that $\sigma_{v^\infty}$ 
is the order-$\alpha_s^0$ contribution in which
order-$v^\infty$ corrections are resummed. The order-$v^\infty$ 
correction $\sigma_{v^\infty}$ is about $-17$\,\% of
$\sigma_{0}$.
We have estimated the order-$v^2$ corrections
from our resummed formula by taking the partial derivative with
respect to $\langle \bm{q}^2 \rangle_{\eta_c}$ numerically.
Our result is $-75.6\,\langle v^2\rangle_{\eta_c}$, which is in good agreement with the
value $-75.5\, \langle v^2\rangle_{\eta_c}$ obtained from the fixed-order calculation
in Ref.~\cite{Sang:2009jc}. We define
$K_{v^2}=\sigma_{v^2}/\sigma_0$ and $K_{v^\infty}=\sigma_{v^\infty}/\sigma_0$,
where $\sigma_{v^2}$ is the cross section that includes the corrections
of order $\alpha_s^0v^2$. The results are $K_{v^2}=0.80$ and $K_{v^\infty}=0.83$
which imply that the contributions of order $v^4$ or higher enhance 
the cross section by about 4\,\%. This is a good signal that the $v$ expansion
converges very well.

The factor $K_{\alpha_s}$ defined by $\sigma_\textrm{NLO} / \sigma_0$ is
$0.77$ for $\mu=m_c$ and $0.82$ for $\mu=2 m_c$,
respectively \cite{Sang:2009jc}.
Taking into account both the NLO corrections in $\alpha_s$ and the relativistic
corrections, and the interference between them, we find the total cross section
$\sigma_\textrm{tot}=51.5^{+13.6}_{-12.8}$ fb for $\mu=m_c$
and $\sigma_\textrm{tot}=55.3^{+14.5}_{-13.7}$ fb for $\mu=2 m_c$,
respectively.
For a luminosity of 1 (10) $\textrm{ab}^{-1}$ at  $B$ (super $B$) factories,
the expected number of events is about $5\times 10^{4}$ ($5\times 10^{5}$).
Since the branching fraction of the $\eta_c$ decay into two photons
is $(1.78\pm0.16)\times 10^{-4}$ \cite{Beringer:1900zz},
$e^+ e^- \to \eta_c + \gamma$ followed by $\eta_c \to \gamma\gamma$
might be observed at Belle II or super $B$ factory, but
the events might not be triggered because no charged particles exist
in the detector. Instead, the $\eta_c\to K\bar{K}\pi$ mode,
whose branching fraction is $(7.2\pm 0.6)$\,\% \cite{Beringer:1900zz},
may be useful to observe the $e^+ e^- \to \eta_c + \gamma$ events.
The analysis of the photon energy spectrum
in $e^+ e^- \to K\bar{K}\pi + \gamma$ will be useful
to detect this channel \cite{Chung:2008km}.
\section{$\bm{e^+e^-\to J/\psi+J/\psi}$
\label{sec:jpsi}}
In this section, we consider the $e^+ e^- \to J/\psi + J/\psi$ process
at $\sqrt{s}=10.58$ GeV. This process proceeds through $e^+ e^-$ annihilation
into two virtual photons because the charge-conjugation parity of $J/\psi$
is $-1$.
\begin{figure}[t]
\begin{center}
{\epsfig{figure=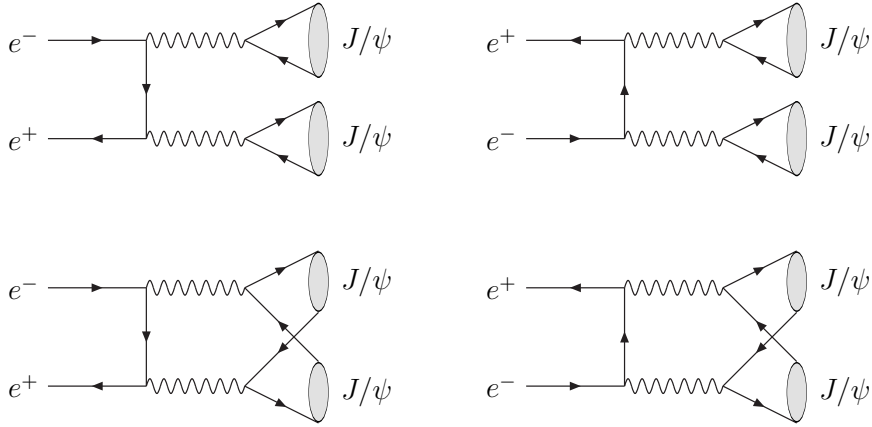,width=0.7\textwidth}}
\end{center}
\vspace{-1.0cm}
\caption{
Feynman diagrams for $e^+ e^- \to J/\psi + J/\psi$.
}
\label{fig:diag2}
\end{figure}
At the LO in $\alpha_s$, there are four Feynman diagrams
as shown in Fig. \ref{fig:diag2}. The first two diagrams are called 
the photon-fragmentation diagrams, where each virtual photon 
evolves into a $J/\psi$. The other two diagrams in the second
row are called the nonfragmentation diagrams, where two virtual photons
evolve into two $J/\psi$ mesons. The invariant mass of the virtual photon
in the photon-fragmentation diagrams is fixed as $m_{J/\psi}$, but 
the typical momentum scale of the virtual photons in the nonfragmentation
diagrams is of order $\sqrt{s}/2$.
\subsection{Amplitude of $\bm{e^+ e^- \to
c\bar{c}({}^3S_1^{[1]})+c\bar{c}({}^3S_1^{[1]})}$}
In order to calculate the relativistic corrections to the double $J/\psi$ 
production in $e^+ e^-$ collisions, we first construct the amplitude for
the production of two color-singlet spin-triplet $S$-wave 
$c\bar{c}({}^3S_1^{[1]})$ pairs in $e^+ e^-$ collisions.
The amplitude for
$e^+(k_2)e^-(k_1)\to c(p_1)\bar{c}(\bar{p}_1) + c(p_2)\bar{c}(\bar{p}_2)$
can be written as
\begin{equation}\label{eq:amplitude2}
\mathcal{M}=\sum_{j=1}^{4}
L^{j \mu\nu} \mathcal{A}^j_{\mu\nu}[c_1\bar{c}_1+c_2\bar{c}_2],
\end{equation}
where $L^{j \mu\nu}$ and $\mathcal{A}^{j}_{\mu\nu}$ are
the leptonic and hadronic tensors for each diagram.
Eventually, each $c_i(p_i)\bar{c}_{i}(\bar{p}_i)$ pair evolves into
a $J/\psi$ meson. 

The amplitude
$\mathcal{A}^j_{\mu\nu}[P_1,q_1,\lambda_1;P_2,q_2,\lambda_2]$
for the production of
two $c\bar{c}({}^3S_1^{[1]})$ pairs from the two virtual photons can be obtained
by projecting the amplitude $\mathcal{A}^j_{\mu\nu}[c_1\bar{c}_1+c_2\bar{c}_2]$
into the color-singlet spin-triplet channel,
where $P_i$, $q_i$, and $\lambda_i$ are the four-momentum of the $c_i\bar{c}_i$
pair, half the relative momentum of $c_i$ and $\bar{c}_i$, and
the polarization vector of the $c_i\bar{c}_i$ pair, respectively, with the
following relations:
\begin{subequations}
\begin{eqnarray}
p_i &=& \frac{P_i}{2} + q_i, \\
\bar{p}_i &=& \frac{P_i}{2} - q_i.
\end{eqnarray}
\end{subequations}
The spin-triplet projector for $c_i\bar{c}_i$ pair, which is valid
for $v^\infty$ corrections, is given by \cite{Bodwin:2002hg}
\begin{equation}
\Pi^i_3(p_i,\bar{p}_i,\lambda_i) =
-\frac{1}{4\sqrt{2}E(q_i)[E(q_i)+m_c]}
(\bar{p}_i\!\!\!\!/-m_c)\,/\!\!\!\epsilon^{\,\ast}_i  (\lambda_i)
[P_i\!\!\!\!\!/+2 E(q_i)](p_i\!\!\!\!/+m_c),
\end{equation}
where $\epsilon_i(\lambda_i)$ is the polarization vector
of the $c_i\bar{c}_i$ pair with the polarization $\lambda_i$
and $E(q_i)$ is the energy of the charm quark or antiquark
in the $c_i\bar{c}_i$ rest frame.
We note that 
the $c_i\bar{c}_i$ pair in
$\mathcal{A}^j_{\mu\nu}[P_1,q_1,\lambda_1;P_2,q_2,\lambda_2]$
is not a pure $S$-wave state.

The $S$-wave contribution
can be extracted by averaging the amplitude over the polar and
azimuthal angles of $\bm{q}_i$, which is the spatial vector of $q_i$,
in the $c_i\bar{c}_i$ rest frame:
\begin{equation}
\label{eq:cccc}%
\mathcal{A}^j_{\mu\nu} 
[(c_1\bar{c}_1)({}^3S_1^{[1]})+(c_2\bar{c}_2)({}^3S_1^{[1]})]
\equiv
\int \frac{d \Omega_1 d \Omega_2}{(4\pi)^2}
\mathcal{A}^j_{\mu\nu} [P_1,q_1,\lambda_1;P_2,q_2,\lambda_2].
\end{equation}
Now the $q_i$ dependence of the quantity in Eq.~(\ref{eq:cccc})
is a function of $\bm{q}_1^2$ and $\bm{q}_2^2$. Here, $\Omega_i$ is 
the solid angle of $\bm{q}_i$ in the $c_i\bar{c}_i$ rest frame.
As we did in the previous section, we do not find the 
short-distance coefficients by expanding Eq. (\ref{eq:cccc}) in terms
of $\bm{q}_i^2$. Instead, we employ the formula (\ref{eq:resum}) 
to obtain the amplitude in which relativistic corrections are resummed:
\begin{eqnarray}
\mathcal{A}^j_{\mu\nu}[J/\psi+J/\psi] &=&
\left.
\left[
\frac{1}{4N_c^2\sqrt{E(q_1)E(q_2)}}
\mathcal{A}^j_{\mu\nu} 
[(c_1\bar{c}_1)({}^3S_1^{[1]})+(c_2\bar{c}_2)({}^3S_1^{[1]})]
\right]
\right|_{\bm{q}_1^2=\bm{q}_2^2=\langle\bm{q}^2\rangle_{J/\psi}}
\nonumber \\
&&\times
\langle O_{0} \rangle_{J/\psi}^{1/2}
\langle O_{0} \rangle_{J/\psi}^{1/2}.
\end{eqnarray}
\subsection{Kinematics}
We first compute the cross section
for $e^+ e^- \to J/\psi + J/\psi$ within the standard NRQCD factorization
approach. Then, as an alternative, we employ the VMD approach for the 
photon-fragmentation diagrams while keeping the standard NRQCD factorization
approach for the nonfragmentation diagrams as in Ref. \cite{Bodwin:2006yd}.
Note that we use the expression $P_i$ for the $c_i\bar{c}_i$ four-momentum
depending on the approach: In the NRQCD approach, we set $P_i^2=[2 E(q_i)]^2$,
while we set $P_i^2=m_{J/\psi}^2$ in the VMD approach.
\subsubsection{Kinematics for the NRQCD factorization approach\label{KinNR}}
In the calculation within the standard NRQCD factorization approach,
the four-momenta of $e^\pm$ and $c_i\bar{c}_i$ pairs 
at the $e^+e^-$ CM frame are given by
\begin{subequations}
\label{eq:kin1}%
\begin{eqnarray}
k_1 &=& \frac{\sqrt{s}}{2}
        \left(1, \sin\theta, 0, \cos\theta\right), 
        \label{NRk1}\\
k_2 &=& \frac{\sqrt{s}}{2}\left(1, -\sin\theta, 0, -\cos\theta\right), 
        \label{NRk2}\\
P_1 &=& \left(E_1, 0, 0, |\bm{P}|\right) ,\\
P_2 &=& \left(E_2, 0, 0,-|\bm{P}|\right) ,
\end{eqnarray}
\end{subequations}
where $E_1 = [s+4 E(q_1)^2-4 E(q_2)^2]/(2\sqrt{s})$,
$E_2 = [s-4 E(q_1)^2+4 E(q_2)^2]/(2\sqrt{s})$, and
$|\bm{P}|=\lambda^{1/2}(s,4 E(q_1)^2, 4 E(q_2)^2)/(2\sqrt{s})$ with
$\lambda(a,b,c)=a^2+b^2+c^2-2ab-2bc-2ca$.
The polarization vectors of the $c_i\bar{c}_i$ pairs are given by
\begin{subequations}
\begin{eqnarray}
\epsilon_1^\ast(0) &=& \frac{1}{\sqrt{E_1^2-|\bm{P}|^2}}
(|\bm{P}|, 0, 0, E_1), \\
\epsilon_2^\ast(0) &=& \frac{1}{\sqrt{E_2^2-|\bm{P}|^2}}
(-|\bm{P}|, 0, 0, E_2), \\
\epsilon_1^\ast(\pm) &=&
 -\epsilon_2^\ast (\mp)
 =
\mp \frac{1}{\sqrt{2}}
(0, 1, \mp i, 0).
\end{eqnarray}
\end{subequations}

In each rest frame of the $c_i\bar{c}_i$ pair,
half the relative momentum $q_i^\ast$ is given by
\begin{equation}
\label{eq:qi}
q_i^\ast = |\bm{q}_i|\left( 0, \sin\theta^\ast_i\cos\phi^\ast_i,
\sin\theta^\ast_i\sin\phi^\ast_i, \cos\theta^\ast_i \right),
\end{equation}
where $\theta^\ast_i$ and $\phi^\ast_i$ are the polar and azimuthal angles
of $\bm{q}_i^\ast$, respectively, and $|\bm{q}_i|$ is the magnitude
of $\bm{q}_i^\ast$ in the $c_i\bar{c}_i$ rest frame.
Boosting Eq. (\ref{eq:qi}) from the $c_i\bar{c}_i$ rest frame
to the $e^+ e^-$ CM frame, one obtains
\begin{subequations}
\begin{eqnarray}
q_1 &=& |\bm{q}_1|\left(
\gamma_1\beta_1\cos\theta_1^\ast, \sin\theta_1^\ast\cos\phi_1^\ast,
\sin\theta_1^\ast\sin\phi_1^\ast, \gamma_1\cos\theta_1^\ast \right),
\\
q_2 &=& |\bm{q}_2|\left(
\gamma_2\beta_2\cos\theta_2^\ast, \sin\theta_2^\ast\cos\phi_2^\ast,
\sin\theta_2^\ast\sin\phi_2^\ast, \gamma_2\cos\theta_2^\ast \right),
\end{eqnarray}
\end{subequations}
where $\gamma_i=E_i/[2E(q_i)]$, $\beta_1=|\bm{P}|/E_1$,
and $\beta_2=-|\bm{P}|/E_2$.
\subsubsection{Kinematics for the VMD approach}
In the calculation within the VMD approach,
the four-momenta and polarization vectors for
the $c_i\bar{c}_i$ pairs at the $e^+e^-$ CM frame are given by
\begin{subequations}
\label{eq:kin2}%
\begin{eqnarray}
P_1^\textrm{fr} &=& (E_1^\textrm{fr}, 0, 0, |\bm{P}^\textrm{fr}|), \\
P_2^\textrm{fr} &=& (E_2^\textrm{fr}, 0, 0,-|\bm{P}^\textrm{fr}|), \\
\epsilon_1^{\textrm{fr}\ast}(0) &=&
\frac{1}{\sqrt{{E_1^\textrm{fr}}^2-|\bm{P}^\textrm{fr}|^2}}
(|\bm{P}^\textrm{fr}|, 0, 0, E_1^\textrm{fr}), \\
\epsilon_2^{\textrm{fr}\ast}(0) &=&
\frac{1}{\sqrt{{E_2^\textrm{fr}}^2-|\bm{P}^\textrm{fr}|^2}}
(-|\bm{P}^\textrm{fr}|, 0, 0, E_2^\textrm{fr}), \\
\epsilon_1^{\textrm{fr}\ast}(\pm) &=&
 -\epsilon_2^{\textrm{fr}\ast} (\mp)
=\mp \frac{1}{\sqrt{2}} (0, 1, \mp i, 0),
\end{eqnarray}
\end{subequations}
where $E_1^\textrm{fr} = E_2^\textrm{fr} =\sqrt{s}/2$ and
$|\bm{P}^\textrm{fr}|=(s-4 m_{J/\psi}^2)^{1/2}/2$.
Here, we use the superscript fr to denote that the variables are 
for the VMD approach. The four-momenta for the $e^\pm$ are the same 
as those in Eqs.~(\ref{NRk1}) and (\ref{NRk2}): $k_i^\textrm{fr}=k_i$.
Because the photon-$J/\psi$ vertex (\ref{eq:gVgamma}) already contains
the relativistic corrections to the photon-fragmentation diagrams,
we do not take the average over the angles of $\bm{q}_1$ and $\bm{q}_2$.
\subsection{ Numerical results for NRQCD prediction \label{sec:novmd}}
Applying the method described in Sec.~\ref{KinNR}, we carry out the
numerical computation of the cross section 
$\sigma[e^+ e^- \to J/\psi + J/\psi]$ at the $B$ factories within
the standard NRQCD factorization approach. We take $\alpha=1/131$ and vary
the charm-quark mass as $m_c=1.4$ and $1.5$ GeV.
The results are summarized in Table \ref{table2} depending on the
input parameters. The LO NRQCD LDME $\langle O_{0} \rangle_{J/\psi}$
and the ratio $\langle \bm{q}^2 \rangle_{J/\psi}$ are listed on the second 
and third columns of Table \ref{table2}, respectively. The values for 
$\langle O_{0} \rangle_{J/\psi}$ and 
$\langle \bm{q}^2 \rangle_{J/\psi}$ at $m_c=1.4$ GeV are taken
from Ref.~\cite{Bodwin:2007fz}. The values at $m_c=1.5$ GeV are obtained 
by applying the same method described in Ref.~\cite{Bodwin:2007fz}.
\begin{table}[t]
\caption{\label{table2}%
The cross sections $\sigma_0$, $\sigma_{v^\infty}$, and
$\sigma_\textrm{tot}$ for $e^+e^-\rightarrow J/\psi+J/\psi$ 
within the standard NRQCD factorization approach in units of fb 
at the scales $\mu=m_c$, $2 m_c$, and $\sqrt{s}/2$
with the input parameters $m_c$,
$\langle O_{0} \rangle_{J/\psi}$, $\langle \bm{q}^2 \rangle_{J/\psi}$,
in units of GeV, GeV$^3$, and GeV$^2$, respectively.
The errors of the cross sections 
reflect the uncertainties of the NRQCD LDMEs only.
}
\begin{ruledtabular}
\begin{tabular}{cccccccc}
\multirow{2}{*}{$m_c$}
&
\multirow{2}{*}{$\langle O_{0} \rangle_{J/\psi}$}
&
\multirow{2}{*}{$\langle \bm{q}^2 \rangle_{J/\psi}$}
&
\multirow{2}{*}{$\sigma_0$}
&
\multirow{2}{*}{$\sigma_{v^\infty}$}
&
\multicolumn{3}{c}{$\sigma_\textrm{tot}$}
\\
&
&
&
&
&
$\mu=m_c$
&
$\mu=2m_c$
&
$\mu=\sqrt{s}/2$
\\
\hline
  $1.4$
&
$0.440^{+0.067}_{-0.055}$
&
$0.441\pm 0.140$
&
$13.39^{+2.88}_{-2.37}$
&
$5.56^{+1.63}_{-1.35}$
&
$-12.74^{+2.51}_{-2.90}$
&
$-2.58^{+0.56}_{-0.60}$
&
$-0.89^{+0.51}_{-0.42}$
\\
  $1.5$
&
$0.436^{+0.065}_{-0.054}$
&
$0.442\pm 0.140$
&
$8.09^{+1.70}_{-1.42}$
&
$3.66^{+1.02}_{-0.85}$
&
$-6.96^{+1.34}_{-1.57}$
&
$-1.36\pm 0.33$
&
$-0.43^{+0.31}_{-0.25}$
\end{tabular}
\end{ruledtabular}
\end{table}

The result for the LO cross section $\sigma_0$ is listed on the third
column of Table \ref{table2}. 
Our NRQCD prediction for $\sigma_0$ is 
greater than a previous result $\sigma_0=6.65\pm 3.02$ fb
in Ref. \cite{Bodwin:2002fk}. This is mainly because our value for
$\langle O_{0} \rangle_{J/\psi}$ is greater than that 
($0.335\pm 0.024$ GeV$^3$) of Ref. \cite{Bodwin:2002fk}.
Our short-distance coefficient at LO in $\alpha_s$
agrees with those in Refs. \cite{Bodwin:2002fk,Gong:2008ce}
and the numerical results also agree once we use the same numerical
value for the LDME.

The cross section $\sigma_{v^\infty}$ in which the relativistic
corrections are resummed to all orders in $v$ is listed
on the fourth column of Table \ref{table2}. The resummed result 
$\sigma_{v^\infty}$ is significantly smaller than the LO prediction 
$\sigma_0$ leading to $K_{v^\infty}=0.42$ ($0.45$) at $m_c=1.4$ ($1.5$)\,GeV.
As we have mentioned, we can extract the order-$v^2$ relativistic corrections
from the resummed formula by varying $\langle \bm{q}^2 \rangle_{J/\psi}$ 
for each $J/\psi$ numerically. The correction is
$-28.4\,\langle v^2 \rangle_{J/\psi}\textrm{\,fb}$
($-17.4\,\langle v^2 \rangle_{J/\psi}\textrm{\,fb}$)
at $m_c=1.4$ ($1.5$)\,GeV for each $J/\psi$, where 
$\langle v^2\rangle_{J/\psi}\equiv\langle \bm{q}^2 \rangle_{J/\psi}/m_c^2$.
Taking into account the fact that there are two $J/\psi$ mesons,
we find that $K_{v^2}=0.05$ ($0.15$) at $m_c=1.4$ ($1.5$)\,GeV.
While the order-$v^2$ corrections are large negative, 
the relativistic corrections of order $v^4$ or
higher significantly enhance the cross section.

The authors of Ref. \cite{Gong:2008ce} computed the cross section 
$\sigma_\textrm{NLO}$ for $e^+ e^- \to J/\psi + J/\psi$ at NLO in $\alpha_s$,
which has a strong dependence on $m_c$ and the renormalization scale $\mu$.
According to the results, the QCD NLO corrections are large negative,
leading to negative cross sections at $\mu=m_c$ for both $m_c=1.4$ and
$1.5$\,GeV. We can extract the factor
$K_{\alpha_s}\equiv \sigma_{\rm NLO}/\sigma_0$
from the results in Table I of Ref. \cite{Gong:2008ce}:
At $m_c=1.4$ ($1.5$)\,GeV,
$K_{\alpha_s}=-0.367$ ($-0.314$), $0.057$ ($0.077$), and
$0.253$ ($0.248$) for $\mu=m_c$, $2m_c$, and $\sqrt{s}/2$, respectively.

Here, we notice that both the QCD NLO and order-$v^2$ relativistic 
corrections are large negative and significantly reduce the cross
section for $e^+ e^- \to J/\psi + J/\psi$. Thus, the interference
between them must be large positive. In addition, the relativistic
corrections of order $v^4$ or higher is large positive. Therefore, 
we expect that the interference contribution can be significant.

A rigorous computation of the interference contribution requires
the information of the factor $K_{\alpha_s}$ for each polarization
of two $J/\psi$ mesons, which is not available yet. Therefore, we
ignore the dependence on the $J/\psi$ polarization and use the formula
(\ref{interference}) for $\sigma_{\rm tot}$ to include the 
QCD NLO corrections, relativistic corrections, and their interference.
Our results for $\sigma_{\rm tot}$ depending on the input parameters 
are given on the last three columns in Table \ref{table2}.\footnote{
At $\mu=m_c$, $K_{\alpha_s}$ is negative and, therefore,
Eq.~(\ref{interference}) is inapplicable to computing $\sigma_\textrm{tot}$.
In this case, we take 
$\sigma_\textrm{tot}=\sigma_{v^\infty}+(K_{\alpha_s}-1)\sigma_0$.}
Every entry for $\sigma_{\rm tot}$ is negative, which is unphysical.
We expect that one can resolve this problem only after including the
corrections of orders $\alpha_s^2$, $\alpha_s v^2$, or higher.
\subsection{ Numerical results with the VMD approach
 \label{sec:vmd}}
In this section, we provide the numerical results for
the cross section for $e^+e^-\to J/\psi+J/\psi$ in which
the photon-fragmentation contribution is computed with the VMD approach.
According to the NRQCD predictions in
Refs.~\cite{Bodwin:2002fk, Bodwin:2002kk}, the nonfragmentation contribution
to $e^+e^-\to J/\psi+J/\psi$ cross section is $0.9$\,\%, and the
interference between the photon-fragmentation and nonfragmentation
contributions is $-15.4$\,\%. This reveals that the photon-fragmentation
contribution dominates in the process. In that limit, the VMD approach
provides a good approximation for the photon-$J/\psi$ vertex in the 
photon-fragmentation diagrams. The coupling in Eq. (\ref{eq:gVgamma})
collects both the QCD NLO and relativistic corrections to the 
photon-$J/\psi$ vertex effectively and, therefore, the theoretical
uncertainties can be greatly reduced~\cite{Bodwin:2007ga}. 
Note that we cannot employ this method to compute the nonfragmentation
diagrams and we apply the conventional NRQCD approach that is described
in the previous section. This hybrid strategy can be justified because
each contribution makes a separate gauge-invariant subset.

In estimating the coupling $g_{J/\psi\gamma}$ in Eq. (\ref{eq:gVgamma}),
we take $\Gamma[J/\psi\to e^+ e^-] =%
(5.55\pm 0.14 \pm 0.02)\,\textrm{keV}$ \cite{Beringer:1900zz}
and $\alpha(\mu=m_{J/\psi}) =1/132.6$, where the scale $\mu$ corresponds
to the momentum transfer at the vertex. In other vertices, we take
$\alpha =1/131$. The resultant numerical value is
$g_{J/\psi\gamma}=0.832\pm 0.010$ GeV$^2$.

\begin{table}[t]
\caption{\label{table3}%
The cross sections $\sigma_0^{\rm fr}$,
$\sigma_{v^\infty}^{\rm fr}$, and
$\sigma_\textrm{tot}^{\rm fr}$ for $e^+e^-\rightarrow J/\psi+J/\psi$
in units of fb, in which the photon-fragmentation contribution is
computed with the VMD approach when
$g_{J/\psi\gamma}=0.832\pm 0.010$ GeV$^2$.
The description on the input parameters
is the same as that in Table~\ref{table2}.
}
\begin{ruledtabular}
\begin{tabular}{cccccccc}
\multirow{2}{*}{$m_c$}
&
\multirow{2}{*}{$\langle O_{0} \rangle_{J/\psi}$}
&
\multirow{2}{*}{$\langle \bm{q}^2 \rangle_{J/\psi}$}
&
\multirow{2}{*}{$\sigma_0^{\rm fr}$}
&
\multirow{2}{*}{$\sigma_{v^\infty}^{\rm fr}$}
&
\multicolumn{3}{c}{$\sigma_\textrm{tot}^{\rm fr}$}
\\
&
&
&
&
&
$\mu=m_c$
&
$\mu=2m_c$
&
$\mu=\sqrt{s}/2$
\\
\hline
  $1.4$
&
$0.440^{+0.067}_{-0.055}$
&
$0.441\pm 0.140$
&
$1.81\pm 0.13$
&
$1.80\pm 0.13$
&
$1.05^{+0.22}_{-0.25}$
&
$1.32^{+0.18}_{-0.20}$
&
$1.39^{+0.17}_{-0.19}$
\\
  $1.5$
&
$0.436^{+0.065}_{-0.054}$
&
$0.442\pm 0.140$
&
$1.85\pm 0.13$
&
$1.84\pm 0.13$
&
$1.35\pm 0.20$
&
$1.49^{+0.17}_{-0.18}$
&
$1.56\pm 0.16$
\end{tabular}
\end{ruledtabular}
\end{table}

Our results for the cross sections
$\sigma_0^{\rm fr}$,
$\sigma_{v^\infty}^{\rm fr}$, and
$\sigma_{\rm tot}^{\rm fr}$ are listed in Table~\ref{table3}. Here, 
the superscript fr indicates that the photon-fragmentation contribution is
computed with the VMD approach while the nonfragmentation diagrams are
computed in the conventional NRQCD factorization approach. The description 
on the input parameters is the same as that in Table~\ref{table2}.
The uncertainties of the cross sections are dominated by the experimental
errors in $\Gamma[J/\psi\to e^+ e^-]$ through $g_{J/\psi\gamma}$.
The LO cross sections $\sigma_0^\textrm{fr}$ at $m_c=1.4$ and $1.5$\,GeV
agree with a previous result $1.69\pm 0.35$ fb in Ref.~\cite{Bodwin:2006yd}
within errors. Sources of this small difference are as follows:
While we use $2 E_i(q_i)$ for the invariant mass of the
$c\bar{c}$ pair for the nonfragmentation diagram,
the authors of Ref.~\cite{Bodwin:2006yd} used
the physical mass $m_{J/\psi}$ to normalize the heavy-quarkonium state.
Another source of the difference is that
we use $\alpha=1/132.6$ to determine $g_{J/\psi}$, whose scale is
taken to be $m_{J/\psi}$, but in Ref. \cite{Bodwin:2006yd},
$\alpha=1/137$ was used.

We notice that the LO cross section $\sigma_0^{\rm fr}$ in Table~\ref{table3}
is significantly smaller than $\sigma_0$ shown in Table~\ref{table2}.
This is because the relativistic corrections to all orders in $v$ are
already included in the fragmentation contribution to $\sigma_0^{\rm fr}$
that was computed with the VMD approach. In addition, 
$\sigma_{v^\infty}^{\rm fr}$ and $\sigma_{v^2}^{\rm fr}$ are essentially
the same as $\sigma_{0}^{\rm fr}$. As a result, 
$K_{v^\infty}^\textrm{fr}\approx K_{v^2}^{\textrm{fr}}\approx 1$.
This indicates that the relativistic correction to the contribution
not coming from photon fragmentation is negligible.

It is important to notice that, in the VMD approach, the QCD and 
relativistic corrections to the photon-$J/\psi$ vertex have been
effectively resummed to all orders.
Therefore, one must be careful to avoid double counting of the QCD NLO
corrections when one combines the NRQCD one-loop corrections and
the relativistic corrections.
In each photon-$J/\psi$ vertex of the photon-fragmentation diagrams,
there is a QCD NLO correction factor $[1-16\alpha_s/(3\pi)]$ at the
cross-section level. Therefore, in order to subtract the amount that
is contained in the NRQCD one-loop corrections in Ref. \cite{Gong:2008ce},
we add $32\alpha_s/(3\pi)$ to the
factor $K_{\alpha_s}$ extracted from the result in Ref. \cite{Gong:2008ce}.
The resultant factor is very close to 1:
$K_{\alpha_s}^\textrm{fr}=0.944$ ($0.939$),  $0.964$ ($0.956$), and
$0.969$ ($0.964$) for $\mu=m_c$, $2m_c$, and $\sqrt{s}/2$, respectively,
at $m_c=1.4$ ($1.5$)\,GeV.

Because most of the QCD and relativistic corrections are already contained
in $\sigma_0^{\rm fr}$, we have found that 
$K_{\alpha_s}^\textrm{fr}\approx 1$ and $K_{v^\infty}^\textrm{fr}\approx 1$.
In this limit, the interference between the QCD and relativistic
corrections is negligible. Therefore, we take
$\sigma_{\rm tot}^{\rm fr}\approx K_{v^\infty}^{\rm fr}\sigma_{0}^{\rm fr}%
+(K_{\alpha_s}^{\rm fr}-1)\sigma_{0}$.
The results for $\sigma_{\rm tot}^{\rm fr}$ are listed on
the last three columns of Table~\ref{table3}.
It is remarkable that every entry for
$\sigma_{\rm tot}^{\rm fr}$ listed in Table~\ref{table3} is positive
in contrast to the corresponding values in Table~\ref{table2}.
This is a strong indication that the corrections of order higher than
$\alpha_s v^0$ may resolve the problem that the NRQCD prediction gives
negative cross section.
In addition, $\sigma_{\rm tot}^{\rm fr}$ is insensitive to the
variation of the scale $\mu$ from $m_c$ to $\sqrt{s}/2$.
Our prediction for $\sigma_{\rm tot}^{\rm fr}$ is consistent with the 
nonobservation of the process at Belle with an upper bound 
$\sigma[e^+ e^- \rightarrow J/\psi + J/\psi] \times \mathcal{B}_{>2}[J/\psi]%
<9.1$\,fb \cite{Abe:2004ww}.
\section{Conclusion
\label{sec:con}}
We have computed the cross sections for the two exclusive production
processes, $e^+ e^- \to \eta_c + \gamma$ and $e^+ e^- \to J/\psi + J/\psi$
at $\sqrt{s}=10.58$ GeV, in which a class of relativistic corrections are 
resummed to all orders in $v$. By including the available QCD corrections
of order $\alpha_s$ and the interference between the QCD and resummed
relativistic corrections, we further improved the corresponding 
theoretical predictions.

The NRQCD prediction for the cross section of
$e^+ e^- \to \eta_c + \gamma$ is listed in Table~\ref{table1}. 
The value $\sigma_{\rm tot}\approx 50$\,fb, that is about
$60$--$70$\,\% of the LO cross section $\sigma_0$,
contains the QCD NLO and resummed relativistic corrections,
and their interference.
In this process, both the QCD NLO and order-$v^2$ relativistic corrections
are negative. The relativistic corrections of order-$v^4$ or higher are 
negligible. The value for $\sigma_\textrm{tot}$ is still large enough to
be measured at $B$ factories or super $B$ factories, for example, through
the $\eta_c\to K\bar{K}\pi$ decay mode.

In the case of $e^+ e^- \to J/\psi + J/\psi$, which is dominated by
the photon fragmentation, we have found that the standard NRQCD prediction
for $\sigma_{\rm tot}$, shown in Table~\ref{table2}, is negative for
every choice of input parameters. As an alternative, we  have
employed the VMD approach to include the QCD and relativistic corrections
to the photon-$J/\psi$ vertex in the photon-fragmentation diagrams.
The VMD result $\sigma_\textrm{tot}^{\rm fr}\approx 1$\,fb, given in 
Table~\ref{table3}, indeed resolves the problem of negative cross section
and the prediction is insensitive to the scale $\mu$.
This is a strong indication that corrections of order higher than
$\alpha_s v^0$ may have significant contributions.
The predicted cross section for $e^+ e^-\to J/\psi+J/\psi$ is
consistent with the nonobservation of the events at Belle.
\begin{acknowledgments}
Y.F. would like to express appreciation to Korea Institute for Advanced Study
(KIAS) for its hospitality while a part of this work was carried out.
We thank KIAS Center for Advanced Computation 
for providing us with the Abacus2 System.
The work of C.Y. was supported in part by Basic Science Research 
Program through the National Research Foundation of Korea (NRF) funded by 
the Ministry of Education Science and Technology (2011-0022996)
and  by NRF Research Grant No. 2012R1A2A1A01006053.
The work of J.L. was supported by the NRF under the
Grant No. 2009-0070667.
\end{acknowledgments}

\end{document}